\documentclass[11pt,a4paper]{article}
\usepackage[margin=25mm]{geometry}
\usepackage{amsmath,amssymb,amsthm,mathtools,booktabs,array}
\usepackage{tikz}
\usepackage[font=small,labelsep=period]{caption}
\usepackage[T1]{fontenc}
\usepackage{lmodern,microtype,xcolor,enumitem,fancyhdr}
\usepackage[colorlinks=true,linkcolor=blue!45!black,citecolor=blue!45!black,urlcolor=blue!45!black]{hyperref}
\hypersetup{pdftitle={A Dichotomy for Real Symmetric Ternary Holant* on Domain Four}}
\setlist{itemsep=3pt,topsep=5pt}
\numberwithin{equation}{section}
\newtheorem{theorem}{Theorem}[section]
\newtheorem*{informaltheorem}{Main theorem (informal)}
\newtheorem{lemma}[theorem]{Lemma}
\newtheorem{proposition}[theorem]{Proposition}
\newtheorem{corollary}[theorem]{Corollary}
\theoremstyle{definition}\newtheorem{definition}[theorem]{Definition}
\theoremstyle{remark}
\newcommand{\R}{\mathbb R}\newcommand{\C}{\mathbb C}
\newcommand{\A}{\mathcal A}\newcommand{\Lsp}{\mathcal L}
\newcommand{\Hol}{\operatorname{Holant}^{*}}\newcommand{\HolC}{\operatorname{Holant}^{*}_{\C}}
\newcommand{\Sym}{\operatorname{Sym}}\newcommand{\GL}{\operatorname{GL}}
\newcommand{\tr}{\operatorname{tr}}\newcommand{\Span}{\operatorname{span}}
\newcommand{\FP}{\mathrm{FP}}
\newcommand{\shP}{\#\mathrm{P}}
\newcommand{\ot}{\otimes}
\newcommand{\affiliation}[1]{\IfBlankTF{#1}{}{\\{\normalsize #1}}}
\newcommand{\email}[1]{\IfBlankTF{#1}{}{\\{\small\href{mailto:#1}{\nolinkurl{#1}}}}}
\title{A Dichotomy for Real Symmetric Ternary Holant$^*$ on Domain Four}
\author{%
  Zhechen Li%
  \affiliation{Ant Group, Beijing, China}%
  \email{lizhechen@pku.edu.cn}%
}
\date{}
\begin{document}
\maketitle
\begin{abstract}
We prove a complexity dichotomy for $\Hol(f)$, where $f$ is a real algebraic symmetric ternary signature on a four-element domain. Liu, Fan and Cai (COCOA 2023, TCS 2025) identified three tractable families and proved that they exhaust the tractable cases for $\{0,1\}$-valued signatures. They conjectured that every real-valued symmetric ternary signature outside these families gives a $\shP$-hard problem. We prove their conjecture for real algebraic weights. We study the real Zariski closure of the group generated by invertible unary slices. On the support of $f$, we obtain an orthogonal splitting, the full general linear group, or an orthogonal complex structure that anticommutes with every slice. In the full-group case we interpolate binary signatures that realize a hard Boolean restriction. In the complex-structure case on domain four, $f$ comes from a complex cubic in two variables. Its factorization gives either a tractable decomposition or an edge gadget that reduces a hard Boolean problem to $\Hol(f)$. In every hard case a fixed finite set of real algebraic unaries suffices. The tractable signatures are exactly the real points of the class of symmetric unitarily decomposable tensors. They are characterized by pairwise commutation of their squared slice matrices.
\end{abstract}

\section{Introduction}\label{sec:intro}

Consider the problem of counting proper edge colorings of a cubic graph with four colors. At each vertex give a triple of distinct colors weight $1$ and every other triple weight $0$. Multiplying the vertex weights and summing over all edge colorings counts each proper coloring exactly once. More generally, a symmetric ternary signature $f$ assigns a weight $f(i,j,k)$ to each unordered triple of colors. In $\Hol(f)$ we may also put any real algebraic unary signature at a vertex of degree one. In particular, we can prescribe the color of its incident edge. Which choices of $f$ allow the resulting sum to be computed in polynomial time?

Cai, Lu and Xia classified Boolean $\Hol$ for symmetric signatures \cite{CLXsymstoc,CLXsym} and then for arbitrary signatures \cite{CLX,CLXjournal}. On three colors the classification extends from one ternary signature to sets of arbitrary arity, where one must decide which signatures remain tractable together (Table~\ref{tab:dichotomies}). For three or more colors, Cai, Guo and Williams \cite{CGWfocs,CGW} classified $\operatorname{Holant}(f)$ for symmetric ternary $f$ invariant under every permutation of the colors, as in the edge-coloring example. Here symmetry refers only to the arguments of $f$, so $f(1,1,2)$ and $f(1,1,3)$ may have different weights.

On four colors, Liu, Fan and Cai \cite{LFCcocoa,LFC} identified three families with polynomial-time algorithms. The algorithms rely on one property of the edge sums. An orthogonal change of basis preserves the sum over the shared color on each edge. In a suitable basis, each of their signatures splits into blocks on mutually orthogonal lines and planes. A nonzero contribution to a connected component uses the same block at every ternary vertex. On a line there is just one product to evaluate. On a plane the signature is proportional to $u^{\ot3}+\bar u^{\ot3}$ with $u=(1,i)^T$. Since $u^Tu=1+i^2=0$, adjacent vertices must choose opposite terms. The choice at one vertex therefore determines the rest. The three families correspond to four lines, a plane and two lines, or two planes.

Liu, Fan and Cai proved that their list is complete when all values of $f$ lie in $\{0,1\}$. The edges still have four colors in that theorem. There are twenty unordered triples of colors, and one chooses which of them have weight $1$. Once arbitrary real algebraic weights are allowed, even a fixed nonempty set of allowed triples gives infinitely many signatures. Liu, Fan and Cai conjectured that every real-valued signature outside their three families gives a $\shP$-hard problem \cite[Open problems]{LFC}.

How might one use a dichotomy on a smaller domain to prove hardness on four colors? Liu, Fan and Cai used binary gadgets to restrict the colors available on an edge \cite{LFCcocoa,LFC}. For instance, if we could put an orthogonal projection of rank three on each edge between ternary vertices, then in an orthonormal basis for its image the edge sums would range over just three colors. The simplest binary gadget attaches a unary $x$ to one edge of $f$. The two remaining edges carry the matrix $A_x$ with entries $A_x(i,j)=\sum_{k=1}^4 f(i,j,k)x_k$. We call this matrix a slice. A search for a slice of rank three can fail. In an example from Liu's thesis \cite[Section~4.3]{LiuThesis}, every slice is either invertible or has rank at most two.

Joining two slices along an edge gives the matrix product $A_xA_y$. In Liu's example, taking two copies of the same slice gives a nonzero multiple of the orthogonal projection onto a plane. The restriction of $f$ to that plane is a hard Boolean signature, as we verify in Appendix~\ref{app:rankthree}. Cai and Ihm used products of binary signatures to act on ternary signatures in their domain-three dichotomy \cite{CI}. Here every matrix in a chain comes from the same symmetric $f$, so $A_xy=A_yx$. We use this identity to study the group generated by the invertible slices (Theorem~\ref{thm:group}). If all slices preserve a common proper nonzero subspace, $f$ splits into two signatures on smaller domains. For the remaining signatures outside the three families, we simulate a hard Boolean problem by interpolation or by an explicit edge gadget.

\subsection{Our Contribution}\label{sec:contribution}

We prove the conjecture of Liu, Fan and Cai for real algebraic weights (Theorem~\ref{thm:main}).

\begin{informaltheorem}
Fix a real algebraic symmetric ternary signature $f$ on four colors. If the matrices $A_x^2$ commute pairwise as $x$ ranges over $\R^4$, then $\Hol(f)$ can be evaluated in polynomial time. Otherwise $\Hol(f)$ is $\shP$-hard, even with a finite set of real algebraic unaries depending only on $f$.
\end{informaltheorem}

The commuting case gives exactly the three families of Liu, Fan and Cai. In every real dimension, commuting squares characterize orthogonal sums of one-dimensional cubes and two-dimensional harmonic blocks (Theorem~\ref{thm:squares}). Section~\ref{sec:certificate} gives finite equations for this test and identifies these cubics as the real symmetric unitarily decomposable tensors \cite{BDHR}.

\begin{center}
\begin{minipage}{\linewidth}
\centering
\captionsetup{type=table,position=top}
\caption{Selected $\Hol$ dichotomies. The weights refer to the symmetric signatures in each row.}
\label{tab:dichotomies}
\small
\begin{tabular*}{\linewidth}{@{\extracolsep{\fill}}lcll@{}}
\toprule
Work & Domain size & Signature family & Weights \\
\midrule
Cai, Lu and Xia \cite{CLX3soda,CLX3} & 3 & One ternary signature & Complex \\
Liu, Fan and Cai \cite{LFCcocoa,LFC}  & 3 & One ternary signature & Real \\
Cai and Ihm \cite{CI}                 & 3 & Any set, arbitrary arities & Real \\
Cai and Ihm \cite{CIcomplex}          & 3 & Any set, arbitrary arities & Complex \\
\addlinespace
Liu, Fan and Cai \cite{LFCcocoa,LFC}  & 4 & One ternary signature & $\{0,1\}$ \\
\textbf{This work}                    & 4 & One ternary signature & Real algebraic \\
\bottomrule
\end{tabular*}
\end{minipage}
\end{center}

\section{Preliminaries and the dichotomy theorem}\label{sec:statement}

\paragraph{Counting problem.} Fix a domain $D=[d]=\{1,\ldots,d\}$ and a family $\mathcal F$ of signatures on $D$. A signature grid $\Omega$ is a finite undirected graph, possibly with loops and parallel edges, with an order on the half-edges incident to each vertex. Each vertex $v$ carries a signature $f_v\in\mathcal F$ of arity $\deg(v)$. An edge labeling $\sigma$ gives a tuple $\sigma_v$ of labels in the prescribed order at $v$. A loop contributes its label twice. The partition function is
\begin{align*}
 Z(\Omega)=\sum_{\sigma\in D^{E(\Omega)}}\prod_{v\in V(\Omega)} f_v(\sigma_v).
\end{align*}
The problem $\operatorname{Holant}(\mathcal F)$ asks for the exact value of $Z(\Omega)$. Write $\Hol(f)$ for the problem with one fixed ternary signature $f$ and all real algebraic unaries. For Boolean source problems, $\HolC(h)$ denotes the variant with complex algebraic unaries. We use polynomial-time Turing reductions, denoted by $\le_T^p$.

\paragraph{Tensor notation.} Let $V=\R^d$ with its standard inner product and standard basis $e_1,\ldots,e_d$. A real symmetric ternary signature $f$ on $[d]$ determines a symmetric tensor in $\Sym^3(V)$, a trilinear form $F$, and a homogeneous cubic $p$ by
\begin{align*}
 F(e_i,e_j,e_k)=f(i,j,k),~~ p(x)=F(x,x,x).
\end{align*}
For $x\in V$, the slice $A_x$ is defined by
\begin{align*}
 \langle A_x y,z\rangle=F(x,y,z).
\end{align*}
Each $A_x$ is real symmetric, depends linearly on $x$, and satisfies $A_xy=A_yx$. Attaching the unary $x$ to one leg of $f$ realizes $A_x$ as a binary signature. Joining such signatures in a chain realizes the ordered product $A_{x_1}\cdots A_{x_k}$. For complex vectors $u,v$, edge contraction uses the bilinear product $u^Tv$. The Hermitian inner product is $u^T\bar v$, where the bar denotes complex conjugation. We write $I$ for the identity on the space in question and $[A,B]=AB-BA$ for the commutator.

For exact evaluation, we allow any fixed real number field containing the entries of $f$ as the input field for unary weights. This includes every fixed finite set of real algebraic unary signatures. The algorithm may use a fixed algebraic extension of the input field. Its running time includes the bit cost of field operations.

The three tractable forms of Liu, Fan and Cai \cite[Theorem~3]{LFC} are built from two kinds of orthogonal blocks. We allow any dimension in the definition since the hardness proof passes to subspaces.

\begin{definition}\label{def:class}
Let $\mathcal C_n$ consist of the cubics on an $n$-dimensional real inner product space that, in some orthonormal coordinates, are orthogonal direct sums of blocks of the forms
\begin{align*}
 a t^3~~\text{and}~~b(s^3-3st^2),~~ a,b\in\R.
\end{align*}
We call the second type a harmonic block. Zero coordinates are allowed. In dimension four the possible block patterns are $1+1+1+1$, $2+1+1$, and $2+2$. We write $f\in\mathcal C_n$ or $F\in\mathcal C_n$ when the associated cubic belongs to $\mathcal C_n$.
\end{definition}

\begin{theorem}[Domain-four dichotomy]\label{thm:main}
Let $f$ be a real algebraic symmetric ternary signature on domain four.
\begin{enumerate}
\item If $f\in\mathcal C_4$, then $\Hol(f)$ has a polynomial-time exact algorithm.
\item If $f\notin\mathcal C_4$, then $\Hol(f)$ is $\shP$-hard, already with a fixed finite set of real algebraic unaries.
\end{enumerate}
Moreover, membership in $\mathcal C_4$ is equivalent to
\begin{align*}
 [A_x^2,A_y^2]=0~~\text{for all }x,y\in\R^4.
\end{align*}
\end{theorem}

Assuming $\FP\ne\shP$, the theorem characterizes polynomial-time solvability. Section~\ref{sec:certificate} proves the matrix criterion in every real dimension.

The hardness proof uses the Boolean Holant$^*$ dichotomy of Cai, Lu and Xia \cite{CLX,CLXsym}, the vanishing Hessian theorem of Gordan and Noether as proved by Bricalli, Favale and Pirola \cite{BFP}, and the low-dimensional representation tables of Nguyen and van der Put \cite{NP}.

\paragraph{Orthogonal summands.} Write $u*v=A_uv$ for the commutative product defined by the tensor.

\begin{lemma}[A common invariant subspace splits the tensor]\label{lem:split}
If a proper nonzero real subspace $W\subset V$ is invariant under all slices, then $F$ is the orthogonal direct sum of its restrictions to $W$ and $W^\perp$,
\begin{align*}
 F=F_W\oplus F_{W^\perp}.
\end{align*}
For algebraic $f$ and $W$ with an algebraic orthonormal basis, the Holant$^*$ problem of each summand reduces to $\Hol(f)$. A fixed finite unary set witnessing hardness for a summand gives such a set for $f$.
\end{lemma}

\begin{proof}
Since each $A_x$ is symmetric, $W^\perp$ is also invariant. For $u\in W$ and $v\in W^\perp$, the product $u*v=A_uv=A_vu$ belongs to both spaces and hence vanishes. By symmetry $F$ has no mixed terms.

Consider a connected instance for the restriction $F_W$. Extend its unaries by zero on $W^\perp$. Each ternary vertex contributes on a single summand, and adjacent ternary vertices must choose the same summand. Thus an extended unary forces the whole component to use $W$. If the component has no unaries, cut one edge and attach the same vector $w_i$ from an orthonormal basis of $W$ to both ends. Sum over $i$. Since $\sum_i w_iw_i^T$ is the orthogonal projection $P_W$ onto $W$, this restores the contraction on $W$ and requires at most $\dim W$ oracle calls. Each component produced by the cut contains a pin. Evaluate unary-only components directly and multiply the component values. For a fixed finite unary set, only the extended unaries and the basis vectors are needed.
\end{proof}

If an algebraic $f$ has a common real invariant subspace of dimension $r$, it also has one with an algebraic orthonormal basis. Indeed, the orthogonal projection onto such a subspace satisfies
\begin{align*}
 P=P^T=P^2,~~ \tr P=r,~~ [P,A_{e_i}]=0~~(1\le i\le d).
\end{align*}
These equations have real algebraic coefficients. Since the real algebraic numbers form a real closed field, the existence of a real solution implies the existence of a real algebraic solution. Orthonormalizing a basis of its image gives the desired basis.

\section{The tractable class}\label{sec:tractable}

Given an orthogonal block decomposition of $f$, we can evaluate a connected component one block at a time. To recognize the tractable class, we need a test in the original coordinates.

\subsection{Evaluation on orthogonal blocks}

\begin{proof}[Proof of the tractability assertion in Theorem~\ref{thm:main}]
For fixed algebraic $f\in\mathcal C_n$, choose the orthonormal coordinates of Definition~\ref{def:class}. For each fixed block pattern, the coefficients of the blocks and the entries of the orthogonal matrix satisfy polynomial equations over the real algebraic numbers. Thus a block decomposition can be chosen over that field. The change of basis preserves edge contractions and transforms each unary by the same orthogonal matrix. In a connected component containing ternary vertices, all such vertices must use the same block. Sum over the at most $n$ choices.

On a one-dimensional block the value is a product of scalar weights. On a harmonic plane the tensor is $\tfrac b2(\beta^{\ot3}+\bar\beta^{\ot3})$, where $\beta=(1,i)^T$. Since $\beta^T\beta=\bar\beta^T\bar\beta=0$ and $\beta^T\bar\beta=2$, adjacent ternary vertices must choose opposite vectors. If the subgraph on ternary vertices is nonbipartite, the block contributes zero. Otherwise this subgraph is connected and has two choices, including when it consists of a single vertex. For each choice take the product of the vertex weights and edge contractions. Evaluate unary-only components directly. The dimension and number field are fixed. Each block contribution is a sum of at most two products, with one factor per vertex or edge. In a fixed field basis the bit length of each product is polynomial in the input size, as is the bit length of the product of the component values. Thus the algorithm runs in polynomial time.
\end{proof}

\subsection{Recognition by slice matrices}\label{sec:certificate}

On a harmonic plane each slice squares to a scalar matrix. The same holds on a one-dimensional block, so all squared slices commute on an orthogonal sum. The converse recovers the blocks from their common eigenspaces.

\begin{theorem}[Commuting squared slices, in every real dimension]\label{thm:squares}
Let $F$ be a real symmetric trilinear form on an $n$-dimensional Euclidean space $V$. Then
\begin{align*}
 F\in\mathcal C_n~~\Longleftrightarrow~~
 [A_x^2,A_y^2]=0~~\text{for all }x,y\in V.
\end{align*}
\end{theorem}

\begin{proof}
Suppose all squares commute. They are real symmetric and can be simultaneously orthogonally diagonalized. Their diagonal entries are quadratic forms in $x$. Group together coordinates with identical quadratic forms, obtaining orthogonal subspaces $W_i$ and pairwise distinct quadratic forms $q_i$ with
\begin{align*}
 A_x^2|_{W_i}=q_i(x)I.
\end{align*}
For generic $x$ the $q_i(x)$ are pairwise distinct. Since $A_x$ commutes with its square, it preserves each eigenspace $W_i$ for these $x$. Each off-diagonal block of $A_x$ is linear in $x$ and vanishes on this dense open set, so it vanishes identically. Lemma~\ref{lem:split} now splits $F$ along the $W_i$.

If $q_i\equiv0$, the symmetric slices vanish on $W_i$. On any other block set $L_x=A_x|_{W_i}$ for $x\in W_i$ and $q=q_i|_{W_i}$. Since there are no mixed terms, $q$ is nonzero. For $x,y\in W_i$, symmetry and $L_xy=L_yx$ give
\begin{align*}
 q(x)\|y\|^2=\|L_xy\|^2=\|L_yx\|^2=q(y)\|x\|^2.
\end{align*}
Thus $q(x)=c\|x\|^2$ with $c>0$. Scale the tensor so $c=1$. Polarizing gives
\begin{align*}
 L_xL_y+L_yL_x=2\langle x,y\rangle I.
\end{align*}
The nonzero cubic has a positive maximum at some unit vector $e$. Lagrange multipliers give $e*e=p(e)e$, while $\|e*e\|=1$, hence $p(e)=1$ and $e*e=e$.

For $y\perp e$, take $x=e$ in the polarized identity and apply it to $e$. Using $L_ye=L_ey$, $L_e^2=I$ and $L_ee=e$, we obtain $0=L_eL_ye+L_yL_ee=y+L_ey$. Thus $L_e$ is $+1$ on $\Span(e)$ and $-1$ on $e^\perp$. If $e^\perp$ contains a nonzero $y$, the invertible operator $L_y$ anticommutes with $L_e$ and exchanges its two eigenspaces. They have equal dimension, so $\dim e^\perp=1$. The block therefore has dimension one or two.

In dimension two, with orthonormal basis $e,v$, one has $e*e=e$, $e*v=-v$, and $\langle v*v,e\rangle=-1$. Since $\|v*v\|=1$, also $v*v=-e$. Therefore the cubic is $s^3-3st^2$, up to its restored scalar. Dimension one gives a cube.

Conversely a one-dimensional cube has scalar squared slices, while a harmonic plane has slices
\begin{align*}
 L_{(s,t)}=b\begin{pmatrix}s&-t\\-t&-s\end{pmatrix},
 ~~ L_{(s,t)}^2=b^2(s^2+t^2)I_2.
\end{align*}
Orthogonal sums of these have commuting squared slices.
\end{proof}

Each block in Definition~\ref{def:class} also satisfies $[A_x,A_y^2]=0$ for all $x,y$. Any matrix commuting with $A_x$ commutes with $A_x^2$, so Theorem~\ref{thm:squares} gives the converse.
\begin{corollary}\label{cor:cubic}
For every real symmetric trilinear form $F$ on an $n$-dimensional Euclidean space $V$,
\begin{align*}
 F\in\mathcal C_n~~\Longleftrightarrow~~ [A_x,A_y^2]=0~~\text{for all }x,y\in V.
\end{align*}
\end{corollary}
Fix an orthonormal basis $e_1,\ldots,e_n$. Set $A_i=A_{e_i}$, $Q_{ii}=A_i^2$, and $Q_{ij}=A_iA_j+A_jA_i$ for $i<j$. Comparing coefficients gives $[A_i,Q_{jk}]=0$ for $1\le i\le n$ and $1\le j\le k\le n$ as an equivalent test. In dimension four these forty commutators are skew-symmetric $4\times4$ matrices. Their entries give at most 240 scalar cubic equations in the twenty independent entries of $f$. The squared-slice criterion instead tests the 45 pairs among the ten $Q_{ij}$.

These criteria describe the real points of the symmetric unitarily decomposable tensor class of Boralevi, Draisma, Horobe\c{t} and Robeva \cite{BDHR}. In a unitary decomposition $\sum_i v_i^{\ot3}$ the vectors $v_i$ are pairwise Hermitian-orthogonal. If the tensor is real, conjugation permutes the rank-one summands by uniqueness \cite[Proposition~7]{BDHR}. For a fixed summand, multiplication by a cube root of unity makes its vector real. For a conjugate pair $v^{\ot3}+\bar v^{\ot3}$, Hermitian orthogonality gives $v^Tv=0$. Thus the real and imaginary parts of $v$ are orthogonal with equal norm, and the pair gives a harmonic cubic on their real span. Distinct conjugation orbits give orthogonal real subspaces. Conversely, the decompositions used in the algorithm give a unitary decomposition of every member of $\mathcal C_n$. The first semi-associativity identity of Boralevi et al.\ \cite[Lemma~20]{BDHR} specializes over $\R$ to $A_xA_yA_z=A_zA_yA_x$. Taking $z=y$ gives the cubic commutator in Corollary~\ref{cor:cubic}.

\section{Structure of the slice group}\label{sec:group}

Lemma~\ref{lem:split} reduces the problem to smaller domains whenever all slices preserve a proper nonzero subspace. To understand what can happen in the remaining case, we study the group generated by invertible slices.

We call $F$ \emph{concise} if the map $x\mapsto A_x$ is injective. Its kernel consists of the vectors $x$ for which $x*y=0$ for every $y\in V$. Thus $F$ is supported on the orthogonal complement of the kernel. Let $\Lsp=\{A_x\mid x\in V\}$ be the space of slices. The \emph{slice group} $G$ is the real Zariski closure in $\GL(V)$ of the subgroup generated by the invertible elements of $\Lsp$. For the following theorem, suppose that $F$ is concise on $V=\R^n$ with $2\le n\le4$, so $\dim\Lsp=n$.

\begin{theorem}[Slice-group alternatives]\label{thm:group}
At least one of the following holds.
\begin{enumerate}
\item $G$ preserves a proper nonzero real subspace, and $F$ splits orthogonally.
\item $G=\GL(V)$.
\item There is an orthogonal operator $J$ on $V$ such that $J^2=-I$ and $A_xJ=-JA_x$ for every $x\in V$. In this case $n$ is even and $F$ is the realification of a complex cubic in $n/2$ variables.
\end{enumerate}
\end{theorem}

\subsection{Invertible slices and the identity component}

The group is defined using only invertible slices. We first show that these are dense in the space of all slices, so that passing to the group preserves the common invariant subspaces.

The Hessian of $p$ at $x$ is $6A_x$. By the theorem of Gordan and Noether \cite[Theorem A]{BFP}, a homogeneous form in at most four variables over an algebraically closed field of characteristic zero with identically zero Hessian determinant is a cone. This means that some nonzero directional derivative vanishes identically. For our cubic, the derivative in direction $u$ at $x$ is $3F(u,x,x)$, which vanishes for all $x$ exactly when $A_u=0$, by polarization. A nonzero complex vector in the kernel of $u\mapsto A_u$ has a nonzero real or imaginary part in the real kernel. Conciseness thus implies that $\det A_x$ is a nonzero polynomial. It is nonzero at some rational $x$, so an invertible slice exists.

The invertible slices form a Zariski dense subset of $\Lsp$. If $G$ preserves a proper nonzero subspace, every slice preserves it by density, and Lemma~\ref{lem:split} gives the first alternative. Assume henceforth that $G$ acts irreducibly on $V$.

Let $H=G^\circ$ be the algebraic identity component. We identify real algebraic groups with their real points and write $H_\C$ for the complexification of $H$. The real points of $H$ are Zariski dense in $H_\C$, since the identity is smooth and has a real neighborhood of full algebraic dimension. Transposition preserves $G$, since it reverses every word in symmetric slices and their inverses. It also preserves $H$. If $A_x$ is invertible, then $tI=A_{tx}A_x^{-1}\in G$ for every nonzero real $t$. The scalar subgroup is algebraically connected and hence lies in $H$.

The set of invertible slices is a nonempty Zariski open subset of the linear space $\Lsp$ and is therefore irreducible. The cosets of $H$ form a finite disjoint closed cover of $G$, so all invertible slices lie in a single coset $HS=SH$, where $S$ is any fixed invertible slice. Let $\A=\Span_\R H$. This is a unital associative algebra closed under transpose, and $S$ normalizes $\A$. For each invertible $A_x$ we have $A_xS^{-1}\in H$. By density,
\begin{align}
 \Lsp\subseteq\A S. \label{eq:coset}
\end{align}
The Jacobson radical of $\A$ is closed under transpose. If $X$ belongs to the radical, then $X^TX$ belongs to it as well. A positive semidefinite nilpotent matrix is zero, so $X^TX=0$ and $X=0$. Thus $\A$ is semisimple.

Its primitive central idempotents are orthogonal projections. Indeed, a central idempotent $E$ commutes with $E^T$, so it is normal and hence self-adjoint. Denote their ranges by $U_1,\ldots,U_r$. Conjugation by $S$ permutes the idempotents, so $SU_i=U_{\pi(i)}$ for some permutation $\pi$. By \eqref{eq:coset}, every slice maps $U_i$ into $U_{\pi(i)}$. For $u\in U_i$ and $v\in U_j$ with $i\ne j$, the equal vectors $A_uv$ and $A_vu$ lie in distinct blocks and hence vanish. For $u,v\in U_i$ and $w\in U_j$ with $j\ne i$, total symmetry gives $\langle u*v,w\rangle=\langle u,v*w\rangle=0$. It follows that $U_i*U_i\subseteq U_i\cap U_{\pi(i)}$. If $\pi(i)\ne i$, then $U_i*V=0$, contrary to conciseness. Thus all slices preserve each $U_i$, and irreducibility of $G$ forces $r=1$.

By the real Wedderburn theorem, $\A$ is isomorphic to the matrix algebra $M_k(\mathbb D)$, where $\mathbb D$ is $\R$, $\C$, or the quaternion algebra $\mathbb H$. If its simple module occurs in $V$ with multiplicity $m$, then
\begin{align*}
 k m\dim_\R\mathbb D=n,~~ k^2\dim_\R\mathbb D=\dim_\R\A\ge n.
\end{align*}
The inequality follows from \eqref{eq:coset}. Since $H$ spans $\A$, it acts reducibly exactly when $m>1$. For $n\le4$ the only possibility with $m>1$ is $n=4$, $k=m=2$ and $\mathbb D=\R$. In this case $\dim_\R\A=\dim\Lsp=4$, so $\Lsp=\A S$. Every $BS$ with $B\in\A$ is then symmetric. Since $S$ is symmetric and invertible, $B^T=S^{-1}BS$ for every $B\in\A$. Conjugation preserves the order of multiplication and transposition reverses it. Thus $\A$ would be commutative, contradicting $\A\simeq M_2(\R)$. This proves that $H$ acts irreducibly on $V$.

\subsection{Real, complex, and quaternionic types}

With $m=1$, the possibilities for $\A$ are $M_2(\R)$ and $\C$ in dimension two, $M_3(\R)$ in dimension three, and $M_4(\R)$, $M_2(\C)$ and $\mathbb H$ in dimension four. In the quaternionic case $\dim_\R\A=\dim\Lsp=4$, so the same transpose argument would force $\mathbb H$ to be commutative.

Suppose $\A\simeq M_{n/2}(\C)$. Its center contains an element $J$ with $J^2=-I$. Transposition sends $J$ to $J$ or $-J$. A real symmetric matrix cannot square to $-I$, so $J^T=-J$ and $J$ is orthogonal. Since $H$ is normal in $G$, conjugation by any $g\in G$ preserves $\A$ and its center, sending $J$ to $\pm J$. All invertible slices have the same sign since they lie in the coset $SH$ and $H$ commutes with $J$. By density, all slices commute with $J$ or all anticommute with $J$.

If they commute, then $F(x,Jy,z)=-F(x,y,Jz)$. By total symmetry,
\begin{align*}
 F(Jx,y,z)=-F(x,Jy,z)=F(x,y,Jz)=-F(Jx,y,z).
\end{align*}
The invertibility of $J$ would give $F=0$. Hence all slices anticommute with $J$. In the complexification $V_\C=V\otimes_\R\C$, the $i$ and $-i$ eigenspaces of $J$ are conjugate and isotropic for the complex bilinear extension of the inner product. Anticommutation gives $F(x,Jy,z)=F(x,y,Jz)$. Thus the complex multilinear extension of $F$ vanishes when two arguments lie in different eigenspaces. Its restrictions to the two eigenspaces are conjugate cubics, giving the realification in the third alternative.

When $F$ is real algebraic, $J$ can also be chosen real algebraic. In a standard orthonormal basis $e_1,\ldots,e_n$, it satisfies
\begin{align*}
 J^T=-J,~~ J^2=-I,~~ A_{e_j}J+JA_{e_j}=0~~(1\le j\le n).
\end{align*}
This system has real algebraic coefficients, so a real solution implies a real algebraic solution.

\subsection{The absolutely real type}

It remains to determine which groups $H$ can have real linear span $\A=M_n(\R)$. The complex linear span of $H_\C$ is $M_n(\C)$, so $H_\C$ acts irreducibly on $\C^n$. Its unipotent radical has a nonzero fixed vector by the theorem of Lie and Kolchin. By normality, the fixed space is $H_\C$-invariant, so it is all of $\C^n$. The action is faithful, hence the unipotent radical is trivial and $H_\C$ is reductive. Schur's lemma and the presence of the scalar subgroup identify the center of its Lie algebra with $\C I$. Therefore
\begin{align*}
 \operatorname{Lie}(H_\C)=\C I\oplus\mathfrak s,
\end{align*}
where $\mathfrak s$ is the semisimple derived Lie algebra. Its action is faithful and irreducible, since the center acts by scalars.

The table lists the faithful irreducible modules of complex semisimple Lie algebras in dimensions two through four. The last column gives the symmetry of an invariant bilinear form when one exists \cite[Section~1.2, pp.~178 and~179]{NP}.
\begin{center}
\begin{tabular}{clll}\toprule
$n$ & $\mathfrak s$ & Module & Invariant form line\\\midrule
2 & $\mathfrak{sl}_2$ & natural & alternating\\
3 & $\mathfrak{sl}_3$ & natural or dual & none\\
3 & $\mathfrak{sl}_2$ & $\Sym^2(\C^2)$ & symmetric\\
4 & $\mathfrak{sl}_4$ & natural or dual & none\\
4 & $\mathfrak{sp}_4$ & natural & alternating\\
4 & $\mathfrak{sl}_2$ & $\Sym^3(\C^2)$ & alternating\\
4 & $\mathfrak{sl}_2\oplus\mathfrak{sl}_2$ & $\C^2\ot\C^2$ & symmetric\\\bottomrule
\end{tabular}
\end{center}
For $\mathfrak s=\mathfrak{sl}_n$, the scalar center gives $\operatorname{Lie}(H_\C)=\mathfrak{gl}_n$. The connected group $\GL_n(\C)$ has the same dimension as its closed subgroup $H_\C$. Hence $H_\C=\GL_n(\C)$ and $G=\GL(V)$.

In each remaining row the module is self-dual. Schur's lemma gives a one-dimensional space of invariant bilinear forms, with the symmetry indicated in the table. The group $G$ normalizes $H$ and its derived Lie algebra, so pullback by $g\in G$ preserves this line. The line is also stable under complex conjugation because $H$ is defined over $\R$. The real or imaginary part of a generator gives a nonzero real generator $B$. Its radical is invariant under $H_\C$, so $B$ is nondegenerate.

For every invertible slice $A_x$, the matrix $A_x^TBA_x$ is a scalar multiple of $B$. Fix an entry $B_{ij}\ne0$ and set $q(x)=(A_x^TBA_x)_{ij}/B_{ij}$. This is a homogeneous quadratic polynomial, and it is nonzero at any invertible slice. By density,
\begin{align}
 A_x^TBA_x=q(x)B~~(x\in V). \label{eq:conformal}
\end{align}
A proper semisimple Lie algebra in the table therefore gives either a symmetric $B$ in dimension at least three or an alternating $B$ in dimension four. The next two lemmas exclude these cases.

\subsection{Two exclusions for commutative products}\label{sec:forms}

Let $*$ be a commutative bilinear product on a finite-dimensional real vector space $V$.

\begin{lemma}[No conformally symplectic commutative multiplication]\label{lem:symplectic}
If $\dim V\ge4$, there is no nondegenerate alternating form $\omega$ on $V$ and nonzero quadratic form $q$ on $V$ satisfying
\begin{align*}
 \omega(x*y,x*z)=q(x)\omega(y,z)~~(x,y,z\in V).
\end{align*}
\end{lemma}

\begin{proof}
Write $T(a,b,c,d)=\omega(a*b,c*d)$ and $b_q(a,c)=q(a+c)-q(a)-q(c)$. The tensor $T$ is symmetric in each pair and antisymmetric under exchange of pairs. Write $X=T(a,b,c,d)$, $Y=T(a,c,b,d)$ and $Z=T(a,d,b,c)$. Polarizing the identity in $x$ gives
\begin{align*}
 X-Z&=b_q(a,c)\omega(b,d),\\
 Y-Z&=b_q(a,b)\omega(c,d),\\
 X-Y&=b_q(a,d)\omega(b,c).
\end{align*}
Subtracting the last two equations from the first gives
\begin{align*}
 b_q(a,c)\omega(b,d)-b_q(a,b)\omega(c,d)-b_q(a,d)\omega(b,c)=0.
\end{align*}
Fix $a$ and put $\alpha=b_q(a,\cdot)$. If $\alpha\ne0$, choose $c$ with $\alpha(c)\ne0$ and take $b,d\in\ker\alpha$. The identity forces $\omega(b,d)=0$. This gives an isotropic hyperplane for a nondegenerate alternating form, which is impossible since $\dim V-1>\dim V/2$. Thus $b_q=0$ and hence $q=0$, a contradiction.
\end{proof}

\begin{lemma}[Commutative composition has dimension at most two]\label{lem:composition}
Suppose $b$ is a nondegenerate symmetric bilinear form on $V$ and $q$ is a nonzero quadratic form satisfying
\begin{align*}
 b(x*y,x*z)=q(x)b(y,z)~~(x,y,z\in V).
\end{align*}
Then $\dim V\le2$.
\end{lemma}

\begin{proof}
Setting $z=y$ and using commutativity gives $q(x)b(y,y)=q(y)b(x,x)$. Thus $q$ is a nonzero scalar multiple of $x\mapsto b(x,x)$. Rescale $b$ so that $b(x,x)=q(x)$. Then $q$ is nondegenerate and $q(x*y)=q(x)q(y)$. Choose $x$ with $q(x)\ne0$. Since $q(x*x)=q(x)^2>0$, a real multiple $a$ of $x*x$ satisfies $q(a)=1$. Left multiplication $L_a$ by $a$ preserves $b$ and is invertible.

Define $u\circ v=(L_a^{-1}u)*(L_a^{-1}v)$. This product is commutative and satisfies $q(u\circ v)=q(u)q(v)$. The element $e=a*a$ has $q(e)=1$ and is a unit, since $e\circ u=a*(L_a^{-1}u)=u$.

Let $M_x$ be left multiplication by $x$ for $\circ$ and let $M_x^\dagger$ be its adjoint with respect to $b$. By polarization, $M_x^\dagger M_x=q(x)I$. For $x$ in the $b$-orthogonal complement $W$ of $e$, apply this identity to $e+tx$ and compare coefficients of $t$. Since $M_e=I$ and $b(e,x)=0$, we get $M_x^\dagger=-M_x$ and hence $M_x^2=-q(x)I$. Applying this to $e$ gives $x\circ x=-q(x)e$. Polarizing on $W$ gives
\begin{align*}
 x\circ y=-b(x,y)e~~(x,y\in W).
\end{align*}
If $W$ is nonzero, its nondegeneracy gives some $x\in W$ with $q(x)\ne0$. The operator $M_x^2$ is invertible, but the displayed identity gives $M_x^2(W)\subseteq\R x$. Thus $\dim W\le1$ and $\dim V\le2$.
\end{proof}

Applying these lemmas to \eqref{eq:conformal} completes the proof of Theorem~\ref{thm:group}.

\section{Hardness}\label{sec:hardness}

The two irreducible cases of Theorem~\ref{thm:group} lead to Boolean signatures in different ways. The full slice group lets us realize a two-dimensional restriction of $f$ by interpolation. In the four-dimensional complex-structure case, the binary cubic is already present, paired with its conjugate.

\subsection{Full-group interpolation}\label{sec:interpolation}

Throughout this subsection, $f$ is a fixed real algebraic symmetric ternary signature on a domain of size $d\ge2$. We write $\GL_d$ for $\GL_d(\R)$. Liu, Fan and Cai used interpolation to replace complex unaries by real ones on domain three \cite[Lemma~9]{LFC}. Their argument extends to any fixed domain. The set of real unaries it uses depends on how often each complex unary occurs in the input. Here each slice requires a unary pin, and we choose the pins once for $f$ by fixing a finite set of generators.

\begin{lemma}[A finite set of generators for interpolation]\label{lem:finitegenerators}
If the slice group of $f$ is $\GL_d$, finitely many invertible slices with real algebraic pins generate a Zariski-dense subgroup of $\GL_d$.
\end{lemma}

\begin{proof}
Some invertible slice $S$ has $S^2$ nonscalar. Otherwise every invertible slice would preserve the standard inner product up to scale, and so would the entire slice group. Since invertibility and the nonscalar condition are open and the slices depend linearly on their pins, we may choose a real algebraic pin for $S$.

Consider finite sets of invertible slices that contain $S$ and $2S$ and have real algebraic pins. Choose such a set for which the Zariski closure $K$ of the generated group has maximal dimension. Adjoining any further such slice $T$ gives a closure $K'$ of the same dimension. Thus $(K')^\circ=K^\circ$, so $T$ normalizes $K^\circ$. Algebraic pins are dense among the pins giving invertible slices, and the normalizer of $K^\circ$ is Zariski closed. It follows that $K^\circ$ is normal in $\GL_d$.

Put $P=S^2$. This matrix is positive definite and nonscalar. To see that $P^t\in K^\circ$ for every real $t$, diagonalize $P$ with positive eigenvalues $\lambda_1,\ldots,\lambda_d$. For any polynomial in the matrix entries, its value at $P^t$ is a sum $\sum_\mu c_\mu\mu^t$, where the distinct positive numbers $\mu$ are products of eigenvalues. If the polynomial vanishes at $P^k$ for every nonnegative integer $k$, the Vandermonde matrix gives $c_\mu=0$ for every $\mu$. Hence the path $t\mapsto P^t$ lies in $K$. It passes through the identity, so it lies in $K^\circ$. Also $(2S)S^{-1}=2I$ belongs to $K$, and its powers are Zariski dense in the scalar subgroup. This subgroup is algebraically connected and lies in $K^\circ$.

Differentiating the path at $t=0$ gives the nonscalar matrix $\log P\in\operatorname{Lie}(K^\circ)$. By normality, this Lie algebra is an ideal of $\mathfrak{gl}_d(\R)$. It contains the scalar matrices, and its image modulo the scalars is a nonzero ideal of the simple Lie algebra $\mathfrak{sl}_d(\R)$. It is therefore all of $\mathfrak{gl}_d(\R)$, and $K^\circ=\GL_d$.
\end{proof}

\paragraph{Interpolation by matrix products.}
Cai, Kowalczyk and Williams used an infinite projective group and projector gadgets to interpolate Boolean unaries \cite{CKWitcs,CKW}. Here we need to interpolate binary signatures. A gadget chain realizes a product of slices, while a word in the group may also use their inverses. We show that the products realized by chains suffice, adapting the argument of Liu and Meng \cite[Appendix K, Theorem K.1]{LM}.

\begin{lemma}[Interpolation of binary signatures]\label{lem:interpolate}
Suppose the slice group of $f$ is $\GL_d$. For every fixed finite set $\mathcal U$ of real algebraic unaries and every fixed finite set $\mathcal B$ of complex algebraic binary signatures, there is a fixed finite set $\mathcal U_0$ of real algebraic unaries such that
\begin{align*}
 \operatorname{Holant}(\{f\}\cup\mathcal U\cup\mathcal B)
 \le_T^p \operatorname{Holant}(\{f\}\cup\mathcal U\cup\mathcal U_0).
\end{align*}
\end{lemma}

\begin{proof}
Fix a finite set $\mathcal S$ of slices as in Lemma~\ref{lem:finitegenerators}, and take their pins for $\mathcal U_0$. Suppose a binary signature $B$ occurs $m\ge1$ times in an instance. Replacing every occurrence of $B$ by a matrix $M$ gives a partition function $Z(M)$ that is homogeneous of degree $m$ in the entries of $M$. Let $v_m(M)$ be the vector of all degree-$m$ monomials in its entries, in a fixed order. Its dimension is
\begin{align*}
 D_m=\binom{m+d^2-1}{d^2-1}.
\end{align*}
Expanding the monomials in the entries of $MA$ gives a linear map $T_A$ with $T_Av_m(M)=v_m(MA)$. For invertible $A$, its inverse is $T_{A^{-1}}$.

Start with any one-letter word $S\in\mathcal S$. Extend each retained word on the right by each generator in $\mathcal S$. Retain an extension if its monomial vector is independent of the vectors already retained. At most $D_m$ words are retained. Every nonempty prefix of a retained word was also retained, so its length is at most $D_m$. The search tests at most $|\mathcal S|D_m$ extensions. At termination the span of the retained vectors is invariant under each $T_A$ for $A\in\mathcal S$. The restriction of $T_A$ to this span is injective, hence surjective. The span is therefore invariant under $T_A^{-1}$ as well. It contains $v_m(I)=T_S^{-1}v_m(S)$ and thus $v_m(g)$ for every $g$ in the group generated by $\mathcal S$.

A linear functional vanishing on this span defines a homogeneous polynomial vanishing on the generated group. By Zariski density it vanishes on $\GL_d$ and hence on the entire matrix space. The monomials are linearly independent as polynomials, so the retained vectors span $\R^{D_m}$. Let $W_1,\ldots,W_{D_m}$ be the matrices of the retained words. Solve for the coefficients $c_j$ in the first equality below. Since $Z$ is a linear functional of the monomial vector, the second equality follows.
\begin{align*}
 v_m(B)=\sum_{j=1}^{D_m}c_j v_m(W_j),~~
 Z(B)=\sum_{j=1}^{D_m}c_j Z(W_j).
\end{align*}
Each $Z(W_j)$ is an oracle value obtained by replacing every occurrence of $B$ by the same nonempty chain of slices. This works for singular or complex $B$ and also on binary cycles. For finitely many binary types, apply the reduction successively.

For fixed $d$, $D_m$ is polynomial in $m$. All calculations take place in a fixed number field containing the entries of the generators and the binary signatures. In a fixed rational basis, multiplication by a generator uses rational coefficients of bounded bit length. The coordinates of a word therefore have bit length linear in its length, and those of its degree-$m$ monomials have polynomial bit length. The rank tests and the linear system can thus be solved in polynomial time.
\end{proof}

\paragraph{Three hard Boolean cubics.}
On the Boolean domain we use $e_1,e_2$ for $0,1$, respectively. We write $h=[h_0,h_1,h_2,h_3]$ for a symmetric ternary signature whose value is $h_j$ on inputs with $j$ ones. Its binary cubic is $h_0s^3+3h_1s^2t+3h_2st^2+h_3t^3$. If this cubic is not the cube of a linear form, $h$ is nondegenerate. For such $h$, the Boolean dichotomy gives $\shP$-hardness of $\HolC(h)$ unless one of two conditions holds \cite[Theorem~2.1]{CLX}. The first requires a pair $(\lambda,\mu)\in\C^2\setminus\{(0,0)\}$ such that
\begin{align}
 \lambda(h_0-h_2)+\mu h_1=0,~~
 \lambda(h_1-h_3)+\mu h_2=0. \label{eq:recurrence}
\end{align}
The second is
\begin{align}
 h_0+h_2=h_1+h_3=0. \label{eq:harmonicrec}
\end{align}
In each hard case, a fixed finite set of complex algebraic unary signatures suffices \cite[Definition~2.2 and footnotes~1 and~4]{CLX}.

\begin{lemma}[A hard Boolean restriction]\label{lem:booleanimage}
If the cubic of $f$ is nonzero and is not a pure cube, there is a real algebraic matrix $R\in\R^{d\times2}$ of rank two such that $\HolC(h)$ is $\shP$-hard for the Boolean signature
\begin{align*}
 h(i,j,k)=F(Re_{i+1},Re_{j+1},Re_{k+1}),~~ i,j,k\in\{0,1\}.
\end{align*}
\end{lemma}

\begin{proof}
Choose a real vector $a$ with $p(a)\ne0$. If $p(sa+tb)$ were a pure cube in $s,t$ for every $b$, comparison of the coefficients of $s^3,s^2t,t^3$ would give
\begin{align*}
 p(b)=F(a,a,b)^3/p(a)^2
\end{align*}
for every $b$. This would make $p$ a pure cube. Hence some linearly independent $a,b$ give a binary restriction that is not a pure cube. The same holds on an open set of pairs, so we may take $a,b$ real algebraic.

The resulting binary cubic has either a double and a simple real root, three distinct real roots, or a real root and a pair of nonreal conjugate roots in the projective line. In the first two cases, choose two distinct real linear factors as coordinates. In the three-root case rescale them so that the third factor is their sum. In the last case take the real factor as $x$ and complete the square in the definite quadratic factor. This gives the following normal forms and signatures, each up to a nonzero scalar.
\begin{align*}
 x^2y &\longleftrightarrow [0,1,0,0],\\
 xy(x+y) &\longleftrightarrow [0,1,1,0],\\
 x(x^2+y^2) &\longleftrightarrow [3,0,1,0].
\end{align*}
For these three signatures the determinant $(h_0-h_2)h_2-h_1(h_1-h_3)$ in \eqref{eq:recurrence} is $-1,-2,2$, respectively. None satisfies \eqref{eq:harmonicrec}, so all three are hard. The roots are algebraic, hence the changes of coordinates may be chosen real algebraic. Composing with the restriction gives the required matrix $R$.
\end{proof}

\begin{corollary}\label{cor:fullhard}
If the cubic of $f$ is not a pure cube and the slice group is $\GL_d$, then $\Hol(f)$ is $\shP$-hard with a fixed finite set of real algebraic unaries.
\end{corollary}

\begin{proof}
Take $R$ from Lemma~\ref{lem:booleanimage}. In an instance of the hard Boolean problem for $h=(R^T)^{\ot3}f$, replace each ternary vertex by $f$. Insert $B=RR^T$ on each edge between ternary vertices and replace each unary $q$ attached to a ternary vertex by $Rq$. Indeed, contracting the shared Boolean index on an internal edge gives $\sum_{i=1}^2 R_{ai}R_{bi}=B_{ab}$, also for a self-loop. On an edge incident to $q$ the contraction gives $(Rq)_a$.

The Boolean problem is hard for finitely many algebraic unary types $q$. For each resulting vector $u=Rq$, choose an algebraic matrix $C$ with $Ce_1=u$. Attaching the real unary $e_1$ to $C$ then realizes $u$. Lemma~\ref{lem:interpolate} eliminates $B$ and all these matrices using finitely many real algebraic unaries. Components containing only unaries can be evaluated directly.
\end{proof}

\subsection{The complex-structure case}\label{sec:holomorphic}

Suppose $J$ is the orthogonal complex structure in Theorem~\ref{thm:group}. In dimension two, the realification of a cubic in one complex variable is a harmonic block after a rotation. In dimension four, choose a real algebraic orthonormal basis with $e_2=Je_1$ and $e_4=Je_3$. Let $E$ be the $4\times2$ matrix with columns
\begin{align*}
 \beta_1=(e_1+i e_2)/\sqrt2,~~
 \beta_2=(e_3+i e_4)/\sqrt2.
\end{align*}
The image $U$ of $E$ and its conjugate $\bar U$ are the two eigenspaces of $J$ over $\C$. Under the bilinear edge contraction they satisfy $E^TE=0$ and $E^T\bar E=I_2$. By Theorem~\ref{thm:group}, all mixed tensor components vanish. Since $f$ is real, we have
\begin{align}
 f=E^{\ot3}P+\bar E^{\ot3}\bar P \label{eq:sector}
\end{align}
for a symmetric binary cubic $P\in\Sym^3(\C^2)$ with algebraic coefficients. A real unary has conjugate components in the two eigenspaces. On a connected instance, two real evaluations will let us recover the contribution from one eigenspace.

\begin{proposition}[Holomorphic branch]\label{prop:holomorphic}
For $f$ of the form \eqref{eq:sector}, a fixed finite set of real algebraic unaries suffices for $\shP$-hardness of $\Hol(f)$ unless $P$ is zero, a pure cube, or a sum $a^{\ot3}+b^{\ot3}$ of cubes of linearly independent Hermitian-orthogonal vectors. In each exceptional case $f\in\mathcal C_4$.
\end{proposition}

\paragraph{Two hard Boolean signatures.}
Let $a,b\in\C^2$ be linearly independent algebraic vectors with $a^Ta\ne0$. Choose a complex algebraic orthogonal change of basis taking $a$ to $(r,0)^T$ and $b$ to $(s,t)^T$ with $rt\ne0$. It preserves edge contractions and maps algebraic unaries to algebraic unaries. If also $a^Tb\ne0$, then $s\ne0$ and the signature of $a^{\ot3}+b^{\ot3}$ becomes
\begin{align*}
 [r^3+s^3,s^2t,st^2,t^3].
\end{align*}
The determinant in \eqref{eq:recurrence} is $r^3st^2\ne0$. The second equality in \eqref{eq:harmonicrec} would give $s^2+t^2=0$, and the first would then give $r^3=0$. Hence the Boolean problem is $\shP$-hard.

With the same $a,b$, define
\begin{align*}
 Q(a,b)=a\ot a\ot b+a\ot b\ot a+b\ot a\ot a.
\end{align*}
In the chosen basis, $Q(a,b)$ has signature $[h_0,h_1,0,0]$ with $h_1=r^2t\ne0$, for every $s$. The second equality in \eqref{eq:recurrence} forces $\lambda=0$ and the first forces $\mu=0$. Also $h_1+h_3\ne0$, so \eqref{eq:harmonicrec} fails. This Boolean problem is $\shP$-hard as well.

\paragraph{An edge gadget for the two eigenspaces.}
For algebraic $w\in\C^2$, attach the unary $u=Ew$ to one leg of $f$. Since $E^TE=0$, only the second term of \eqref{eq:sector} contributes, giving the binary matrix $\bar E B\bar E^T$ with
\begin{align*}
 B_{ij}=\sum_{k=1}^2\bar P_{ijk}w_k.
\end{align*}
Suppose that $B$ is invertible. Diagonalizing this symmetric form by congruence and taking square roots of its nonzero diagonal entries gives $B=S^TS$ for an invertible complex algebraic matrix $S$.

Consider a connected Boolean instance with ternary signature $S^{\ot3}P$ and at least one ternary vertex. Replace every ternary vertex by $f$ and insert the binary gadget on every edge, including edges incident to unaries. The gadget pairs only with the $U$ component at either endpoint. Each edge between ternary vertices contributes $B_{ij}=\sum_{k=1}^2 S_{ki}S_{kj}$ in the basis given by $E$. For a Boolean unary $q$, use the endpoint vector $v=ES^{-1}q$. The edge then contributes $BS^{-1}q=S^Tq$. Absorbing the factors of $S$ into the ternary vertices gives precisely the value $Z$ of the Boolean instance.

Replace the gadget pin $u$ by $u+\bar u$ and each endpoint vector $v$ by $v+\bar v$. All these unaries are real algebraic. Each gadget now contributes either its original component in $\bar U\ot\bar U$ or its conjugate in $U\ot U$. Figure~\ref{fig:edge-gadget} shows the two compatible choices on an edge between ternary vertices. In either case the endpoints must choose the same eigenspace, so connectivity leaves just the contributions $Z$ and $\bar Z$.

\begin{center}
\begin{minipage}{\linewidth}
\captionsetup{type=figure}
\centering
\begin{tikzpicture}[
    x=1cm, y=1cm, font=\small,
    wire/.style={draw, line width=.65pt},
    vertex/.style={circle, draw, line width=.65pt, fill=white,
                   inner sep=0pt, minimum size=5.5mm},
    gadget/.style={draw, dashed, line width=.55pt, rounded corners=2pt},
    component/.style={gadget, fill=white, minimum width=18mm,
                      minimum height=5.8mm, inner sep=1.5pt}
  ]
  \node[vertex] (left)   at (-1.65,0) {$f$};
  \node[vertex] (middle) at (0,0)     {$f$};
  \node[vertex] (right)  at (1.65,0)  {$f$};
  \node[draw, rounded corners=1pt, inner xsep=4pt, inner ysep=2pt]
    (pin) at (0,1) {$u+\bar u$};
  \draw[wire] (left) -- (middle) -- (right);
  \draw[wire] (middle) -- (pin);
  \draw[wire] (left) -- (-2.2,.44) (left) -- (-2.2,-.44);
  \draw[wire] (right) -- (2.2,.44) (right) -- (2.2,-.44);
  \draw[gadget] (-.75,-.4) rectangle (.75,1.4);
  \node at (0,-.87) {(a) The real edge gadget};

  \begin{scope}[xshift=6cm]
    \node[vertex] (lu) at (-1.6,.96) {$U$};
    \node[component] (bb) at (0,.96) {$\bar U\otimes\bar U$};
    \node[vertex] (ru) at (1.6,.96) {$U$};
    \draw[wire] (lu) -- (bb) -- (ru);

    \node[vertex] (lb) at (-1.6,-.06) {$\bar U$};
    \node[component] (uu) at (0,-.06) {$U\otimes U$};
    \node[vertex] (rb) at (1.6,-.06) {$\bar U$};
    \draw[wire] (lb) -- (uu) -- (rb);
    \node at (0,-.87) {(b) Compatible eigenspaces};
  \end{scope}
\end{tikzpicture}
\caption{A real unary pin synchronizes the endpoints of an edge. The dashed box in (a) has the two components shown in (b). The top row comes from $u$ and the bottom row from $\bar u$.}
\label{fig:edge-gadget}
\end{minipage}
\end{center}

For a second query, use $iu-i\bar u$ as the pin on one edge gadget. This multiplies the first contribution by $i$ and the second by $-i$. The two oracle values $Z_0,Z_1$ therefore satisfy
\begin{align*}
 Z_0=Z+\bar Z,~~ Z_1=iZ-i\bar Z,~~ Z=(Z_0-iZ_1)/2.
\end{align*}
Evaluate distinct connected components separately and multiply their values. Components containing only unaries can be evaluated directly. For fixed $w,S$ and a finite set of Boolean unary types $q$, the two gadget pins and all endpoint unaries belong to a fixed finite real algebraic set.

\paragraph{Three distinct roots.}
A complex binary cubic with three distinct projective roots is linearly equivalent up to scale to $x^3+y^3$, since a projective linear transformation can take any ordered triple of distinct points of $\mathbb P^1(\C)$ to any other. For algebraic $P$, its roots and the transformation can be chosen algebraic. Absorbing the scale into the linear forms gives $P=a^{\ot3}+b^{\ot3}$ with independent algebraic vectors $a,b$.

Put $c=a^T\bar b$ and write $\|a\|^2=a^T\bar a$ for the squared Hermitian norm. For any positive integer $t$, choose $w$ with $\bar a^Tw=1$ and $\bar b^Tw=t$. Such an algebraic $w$ exists because $\bar a,\bar b$ are independent. Then
\begin{align*}
 B=\bar a\bar a^T+t\bar b\bar b^T=S^TS,~~
 S=\begin{pmatrix}\bar a^T\\ \sqrt t\,\bar b^T\end{pmatrix}.
\end{align*}
Here $S$ is invertible. The transformed cubic is the sum of the cubes of $Sa$ and $Sb$. If $c\ne0$, the two expressions
\begin{align*}
 (Sa)^T(Sa)=\|a\|^4+t c^2,~~
 (Sa)^T(Sb)=\|a\|^2\bar c+t c\|b\|^2
\end{align*}
are nonzero polynomials of degree at most one in $t$. Some $t\in\{1,2,3\}$ avoids both roots. The first Boolean case now gives hardness via the real edge gadget.

If $c=0$, then $a$ and $b$ are Hermitian-orthogonal. A unitary change of coordinates on $\C^2$ takes them to the two coordinate axes and induces a real orthogonal change on $\R^4$. Each coordinate cube realifies to a harmonic cubic on its plane after a rotation. The two planes are orthogonal, so $f\in\mathcal C_4$.

\paragraph{A double root and the remaining cases.}
If $P$ has a double root and a simple root, its factorization into linear forms gives $P=Q(a,b)$ for independent algebraic $a,b$, with the scalar absorbed into $b$. Choose $w$ with $\bar a^Tw=1$ and $\bar b^Tw=t$. Then
\begin{align*}
 B=t\bar a\bar a^T+\bar a\bar b^T+\bar b\bar a^T
   =\begin{pmatrix}\bar a&\bar b\end{pmatrix}
    \begin{pmatrix}t&1\\1&0\end{pmatrix}
    \begin{pmatrix}\bar a&\bar b\end{pmatrix}^{T}.
\end{align*}
All three factors are invertible for every $t$. Moreover,
\begin{align*}
 a^TBa=t\|a\|^4+2\|a\|^2 a^T\bar b
\end{align*}
is nonzero for at least one of $t=0,1$. Choose such a $t$ and factor $B=S^TS$ as above. Then $(Sa)^T(Sa)=a^TBa\ne0$. Since $S^{\ot3}P=Q(Sa,Sb)$, the second Boolean case gives hardness via the same gadget.

A nonzero cubic with a triple root is a pure cube and realifies to one harmonic block. The zero cubic gives $f=0$. This completes the proof of Proposition~\ref{prop:holomorphic}.

\subsection{Completion of the dichotomy}\label{sec:assembly}

\begin{proof}[Completion of the proof of Theorem~\ref{thm:main}]
Section~\ref{sec:tractable} proves tractability and gives the recognition criterion. We prove by induction on $n\le4$ that every $f$ outside $\mathcal C_n$ is hard for a fixed finite set of real algebraic unaries. The zero tensor belongs to $\mathcal C_n$, and every cubic in one variable belongs to $\mathcal C_1$.

Suppose $2\le n\le4$ and $f\ne0$. If $f$ is nonconcise, split off the kernel of $x\mapsto A_x$. If $f$ is concise and its slice group acts reducibly, choose a proper nonzero invariant subspace with an algebraic orthonormal basis as in Section~\ref{sec:statement}. In both cases Lemma~\ref{lem:split} gives two orthogonal summands of smaller dimension. If both have tractable block decompositions, these combine to give one for $f$. Otherwise the induction hypothesis gives a finite hardness witness for a summand, and the lemma lifts it to $f$.

It remains to consider a concise tensor whose slice group acts irreducibly. In the full-group case, conciseness rules out a pure cube, and Corollary~\ref{cor:fullhard} gives the required witness. In the complex-structure case, the dimension is two or four. The two-dimensional case gives one harmonic block. In dimension four, Proposition~\ref{prop:holomorphic} gives either a tractable block decomposition or a finite real algebraic unary witness.
\end{proof}

Interpolation for the full slice group and the slice-matrix recognition criterion hold in every fixed dimension. Section~\ref{sec:forms} excludes a nondegenerate symmetric form up to scale in dimensions at least three and an alternating one in dimensions at least four. Beyond dimension four, the structural argument must also account for concise cubics with identically vanishing Hessian determinant, further irreducible representations, and complex cubics in more than two variables.

\pdfbookmark[1]{\refname}{references}
\bibliographystyle{plain}
\bibliography{ref}

\clearpage
\appendix
\section{A concrete obstruction to a rank-three-slice strategy}\label{app:rankthree}

In an example from Liu's thesis \cite[Section~4.3, p.~79]{LiuThesis}, the four coordinate slices $A,B,C,D$, in that order, are
\begin{align*}
 &A=\begin{pmatrix}0&1/2&1&1\\1/2&0&0&0\\1&0&0&0\\1&0&0&0\end{pmatrix},~~
 B=\operatorname{diag}(1/2,-1/8,1,1),\\
 &C=\begin{pmatrix}1&0&0&0\\0&0&1&0\\0&1&5&1\\0&0&1&1\end{pmatrix},~~
 D=\begin{pmatrix}1&0&0&0\\0&0&0&1\\0&0&1&1\\0&1&1&5\end{pmatrix}.
\end{align*}
For real $x,y,z,w$, their linear combination has determinant
\begin{align*}
 \det(xA+yB+zC+wD)=-\frac{(2w+y+2z)^4}{16}.
\end{align*}
On the singular hyperplane $y=-2w-2z$, set
\begin{align*}
 r&=(0,1/2,1,1)^T,\\
 u&=(x,(w+z)/4,(3z-w)/2,(3w-z)/2)^T.
\end{align*}
Substitution gives $xA+yB+zC+wD=ru^T+ur^T$, which has rank at most two. Thus no slice has rank three.

The square of $A$ nevertheless yields a projection. Let
\begin{align*}
 v=(0,1/3,2/3,2/3)^T,~~ W=\Span\{e_1,v\}.
\end{align*}
The vectors $e_1,v$ are orthonormal, and $A^2=(9/4)P_W$. A chain of two $A$ gadgets with one pin scaled by $4/9$ realizes $P_W$. Inserting this chain on every edge between ternary vertices and embedding the Boolean unaries in $W$ gives the restriction to $W$. In the basis $e_1,v$, its signature is $[0,a,0,b]$ with $a=3/2$ and $b=45/8$. Its coordinate slices are
\begin{align*}
 L_1=\begin{pmatrix}0&a\\a&0\end{pmatrix},~~ L_2=\operatorname{diag}(a,b).
\end{align*}
The upper right entry of $[L_1,L_2^2]$ is $a(b^2-a^2)\ne0$. By Corollary~\ref{cor:cubic} and the dimension-two dichotomy proved in Section~\ref{sec:assembly}, the restriction is hard for a fixed finite set of real algebraic unaries. The projection gadget lifts this witness to the original signature.
\end{document}